\documentclass[12pt]{article}
\usepackage{geometry} % see geometry.pdf on how to lay out the page. There's lots.
\usepackage{graphicx}
\usepackage{amsmath}        % For < align > environment and < \substack >

\usepackage[perpage,symbol]{footmisc}	% footnote markers are symbols

\usepackage{natbib}

\def\cyte#1{\citep{#1}}
\def\mycaption#1#2{\caption{\textbf{#1} #2}}

\def\strutdepth{\dp\strutbox}
\def\marginstar{\strut\vadjust{\kern-\strutdepth\specialstar}}
\def\specialstar{\vtop to \strutdepth{\baselineskip\strutdepth\vss\llap{$\clubsuit$\ \ \ \ }\null}}

\def\revised#1{#1} 
\def\forWsmallP{\ \ \ \ \ \bigg[\textrm{for}\ w\ge\frac{10}{EP}\ \textrm{and small}\ P\bigg]}
\def\Lww{\overline{L(w)}}

\def\appendixEqn#1{}

\usepackage[right]{lineno}
\usepackage{color}

\definecolor{mygray}{gray}{0.7}

\title{Biotechnology and the lifetime of technical civilizations}
\author{John G.\ Sotos, MD\footnote{Air Division, Joint Forces Headquarters, California National Guard, Sacramento, CA 95826. The views expressed are those of the author and do not necessarily reflect the official policy or position of the California Military Department, the Air Force, the Department of Defense, or the U.S. Government.}}
\date{November 28, 2018}

\def\we{I}
\def\OurModel{The model}
\def\private#1{ $<$#1$>$}
\def\private#1{}

\begin{document}

\maketitle

\centerline{\textbf{Abstract}}
\medskip

\noindent The number of people able to end Earth's technical civilization has heretofore been small.  Emerging dual-use technologies, such as biotechnology, may give similar power to thousands or millions of individuals.   To quantitatively investigate the ramifications of such a marked shift on the survival of both terrestrial and extraterrestrial technical civilizations, this paper presents a two-parameter model for civilizational lifespans, i.e. the quantity $L$ in Drake's equation for the number of communicating extraterrestrial civilizations. 
One parameter characterizes the population lethality of a civilization's biotechnology and the other characterizes the civilization's psychosociology.  $L$ is demonstrated to be less than the inverse of the product of these two parameters.  Using empiric data from PubMed to inform the biotechnology parameter, the model predicts human civilization's median survival time as decades to centuries, even with optimistic psychosociological parameter values, thereby positioning biotechnology as a proximate threat to human civilization.  For an ensemble of civilizations having some median calculated survival time, the model predicts that, after 80 times that duration, only one in $10^{24}$ civilizations will survive -- a tempo and degree of winnowing compatible with Hanson's ``Great Filter.''  Thus, assuming that civilizations universally develop advanced biotechnology, before they become vigorous interstellar colonizers, the model provides a resolution to the Fermi paradox.

\vfill\break

%%%%%%%%%%%%%%%%%%%%%%%%%%%%%%%%%%%%%%%%%%%%%%%%%%%%%%%%%%

\section{Introduction}

In 1961 Drake introduced a multi-parameter equation to estimate the number of civilizations in the galaxy capable of interstellar communication\footnote{For brevity, ``civilization'' in this paper refers to a civilization capable of interstellar communication, and the ``lifespan'' or ``lifetime'' of a civilization is the span of time during which it is able to communicate.  Thus, the ``death,'' ``silencing,'' or ``ending'' of a civilization are synonymous.} \cyte{drake1961}.  Soon after, von Hoerner, Shklovskii, and Sagan \cyte{vonhoerner1961} \cyte{shklovskiiSagan}
concluded that the equation's precision depended principally on its parameter $L$ -- the mean lifetime of a communicating civilization -- because $L$'s value was uncertain over several orders of magnitude.  While subsequent advances in astrophysics have  improved the precision of several parameters in the Drake equation \cyte{burchell_2006} \cyte{frank2016} \cyte{vakoch2015}, $L$ remains highly uncertain \cyte{cyclops1971} \cyte{ambartsumian1973} \cyte{billingham1979} \cyte{duncan1991} \cyte{schenkel1999} \cyte{kompanichenko2000} \cyte{rubin2001} \cyte{forgan2009} \cyte{maccone2010}.  

The apparent absence of communicating civilizations \cyte{webb2015} in our planet-rich galaxy \cyte{cassan2012} underscores the possibility that such civilizations have short $L$ \cyte{webb2015} \cyte{bostrom2001}, potentially due to factors exogenous to the civilization (e.g., nearby super\-novae) and/or endogenous to the civilization (e.g., self-destruction).

On Earth, control of endogenous factors that could destroy civilization -- namely, Malthusian resource exhaustion, nuclear weapons, and environmental corruption -- has until now rested with the very few persons who command large nuclear arsenals or steer the largest national economies.  However, emerging technologies could change this.
For example, biotechnology \cyte{pcast2016} and nanotechnology \cyte{drexler1987} offer the prospect of self-replicating elements able to spread autonomously and calamitously worldwide, at low cost and without heavy industrial machinery.  Ultimately, thousands of individuals -- having varying levels of impulse control -- could wield such technologies.

Intuition suggests danger rises as potentially civilization-ending technology  (``CE technology'') becomes more widely distributed, but quantitative analyses of this effect in the context of Drake's $L$ are rare.  At the extreme of technology diffusion, Cooper \cyte{cooper2013} modeled an entire population of 10$^{10}$ individuals (growing at 2\%\ annually), each with a 10$^{-7}$ annual probability of unleashing a biological agent causing 50\%\ mortality (with 25\%\ standard deviation).  He found a mean span of $L$=8000 years before extinction, defined as a population less than 4000.

This article generalizes Cooper's work. It develops a simple two-parameter mathematical model for $L$ that applies to most scenarios of disseminated CE technology and is mathematically indifferent to specific CE technologies. For reasons summarized below, however, biotechnology may be regarded as a universal CE technology. 

\section{Biotechnology's potential to end civilizations}

On Earth, microbial pandemics have ended non-technical civilizations \cyte{mcneill1976}.
Antimicrobial drugs mitigate such risks only partially.  Advisors to the President of the United States have already warned that biotechnology's rapid progress may soon make possible engineered microorganisms that hold ``serious potential for destructive use by both states and technically-competent individuals with access to modern laboratory facilities'' \cyte{pcast2016}.  Indeed, small research groups engineered proof-of-principle demonstrations years ago \cyte{jackson2001} %virulence
\cyte{herfst2012} \cyte{imai2012}, % transmissibility
while recent history provides a precedent not only for a laboratory-preserved organism causing a worldwide pandemic\footnote{This pandemic miserably sickened the author in early 1978.} \cyte{wertheim2010} \cyte{rozo2015}, but also for the organism's descendants circulating for 30 years in the global population \cyte{zimmer2009}. Looking forward, medical research initiatives such as the Cancer Moonshot \revised{\cyte{cancermoonshot2018}} may, if successful, seed thousands of hospitals with exquisitely targetable cell-killing biotechnology that could, in principle, be adapted and aimed at any genetically defined target, not just cancer cells.

Any \revised{technically-capable} intelligence produced by evolution likely shares this susceptibility.   ``Genetic'' processes, defined here as those that pass information to build a succeeding generation or direct the self's use of sustaining energy, are required for evolution \cyte{Farnsworth2013}.  Assuming that no process can be perfect, imperfections in genetic processes equate to ``genetic diseases,'' and will spur any intelligence having self-preservation drives to develop genetic manipulation technology to ameliorate those diseases.  \revised{Given this motivation to alter genetic processes, plus the biological certainty that genetic processes respond to environmental inputs (e.g.\ food shortages), plus a general technical capacity to control environments ever more precisely, the eventual appearance of biotechnology may be expected.} Cooper \cyte{cooper2013} expects that civilizations will typically develop biotechnology and spaceflight approximately simultaneously.

Biotechnology is inescapably threatening because it is inherently dual-use \cyte{watson2018}: curing genetic disease enables causing genetic disease. Cooper \cyte{cooper2013} uses Cohen's theorem \cyte{cohen1987} to assert that, under any reasonable model of computing (applied here to bio-molecular computing), no algorithm (``medical treatment'') can stop every possible piece of invasive self-replicating software.  Whether Cohen's theorem strictly applies or not, the truism that defensive technology generally lags offensive is relevant.

Of course, any civilization can walk away from any technology. But, because other widely available technologies with civilization-ending potential, e.g., nanotechnology, lack the \textsl{a priori} universal desirability of biotechnology, only biotechnology will herein be further discussed. 

%%%%%%%%%%%%%%%%%%%%%%%%%%%%%%%%%%%%%%%%%%%%%%%%%%%%%%%%%%
\section{Model and Results}

The \revised{baseline} model assumes that all communicating technical civilizations either continue communicating forever, or go silent involuntarily due to some action arising within each civilization.  Two parameters model the lifespan of such civilizations: $E$, the number of entities (individuals, coalitions, nation-states, etc.) in the civilization who control a means to end civilization (i.e., render it uncommunicative), and $P$, the uniform probability per annum per entity that an entity will trigger its civilization-ending means.  Entities act independently, and civilization is assumed to end with the first trigger.

The simplest model for the probability, $C(y)$, that the civilization will still be communicative after $y$ years, under constant $E$ and $P$, is:

\begin{equation}            % Creates equation number
C(y) = (1-P)^{(Ey)}
\label{eqn:defCy}
\end{equation}

\noindent Solving Equation~\ref{eqn:defCy} for $y$:

\begin{equation}            % Creates equation number
y = \frac{ln\ C(y)}{E \  ln( 1-P )}
\label{eqn:defY}
\end{equation}

Borrowing the abbreviation $LD_{50}$ from pharmacology, where it indicates the median lethal dose of a substance, it is here re-conceptualized as ``lethal duration 50'' to indicate the number of years, under a given $E$ and $P$, before civilization's accumulated probability of being uncommunicative, $1-C(y)$, is 50\%. Substituting $C(y)=1-0.50$ into Equation~2 yields:
\begin{equation}            % Creates equation number
LD_{50} = \frac{ln(1-.50)}{E \ ln(1-P)} 
\label{eqn:defLDFifty}
\end{equation}

Similarly, the number of years before civilization has a 5\% chance of becoming uncommunicative is:

$$LD_{05} = \frac{ln(1-.05)}{E \ ln(1-P)} \approx (0.074\ LD_{50}) \approx \frac{LD_{50}}{13.5}$$

\begin{figure}[!ht] %%%%%%%% Figure 1
    \hrule
    \centerline{\includegraphics[scale=0.39]{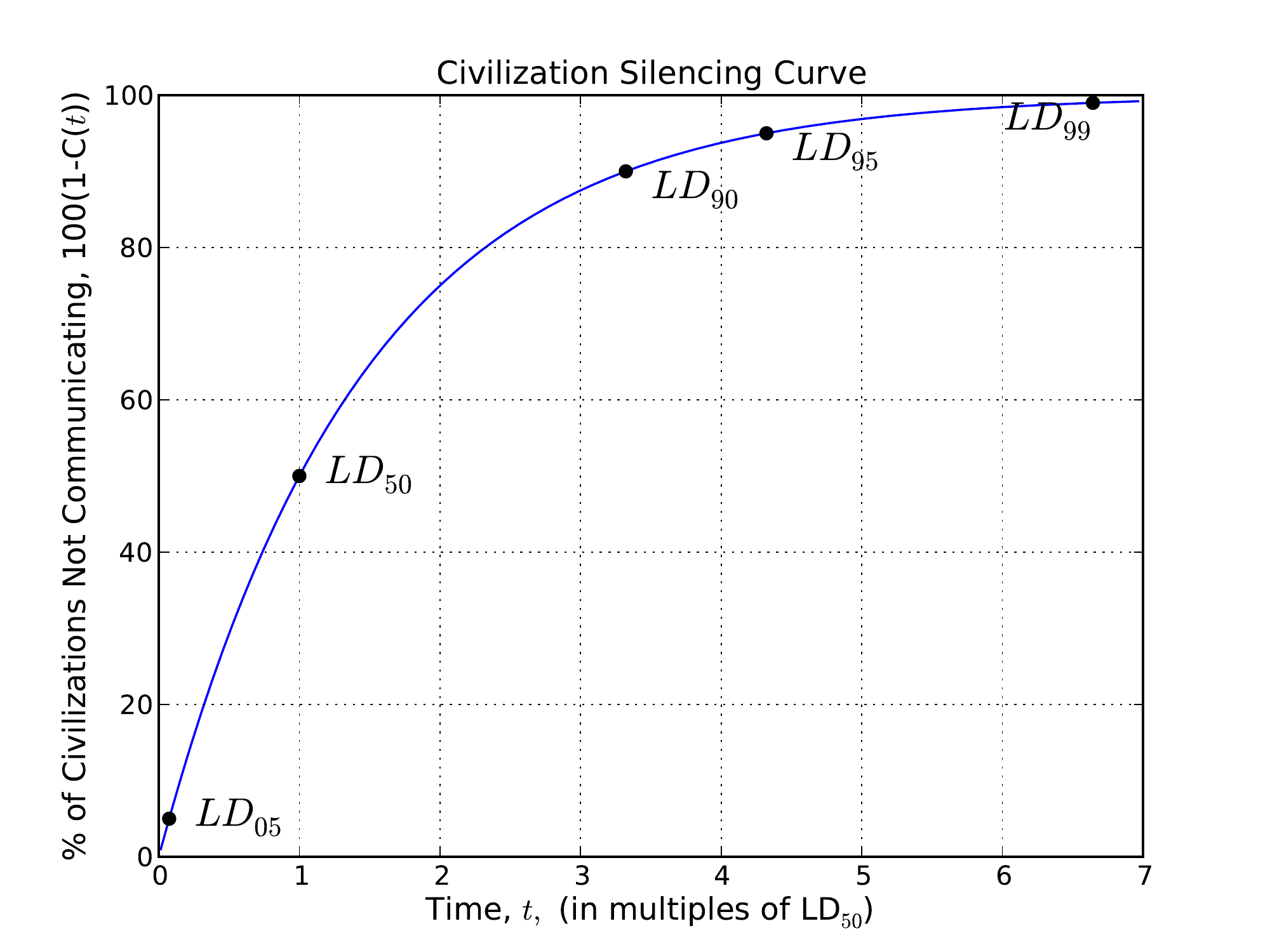}\includegraphics[scale=0.39]{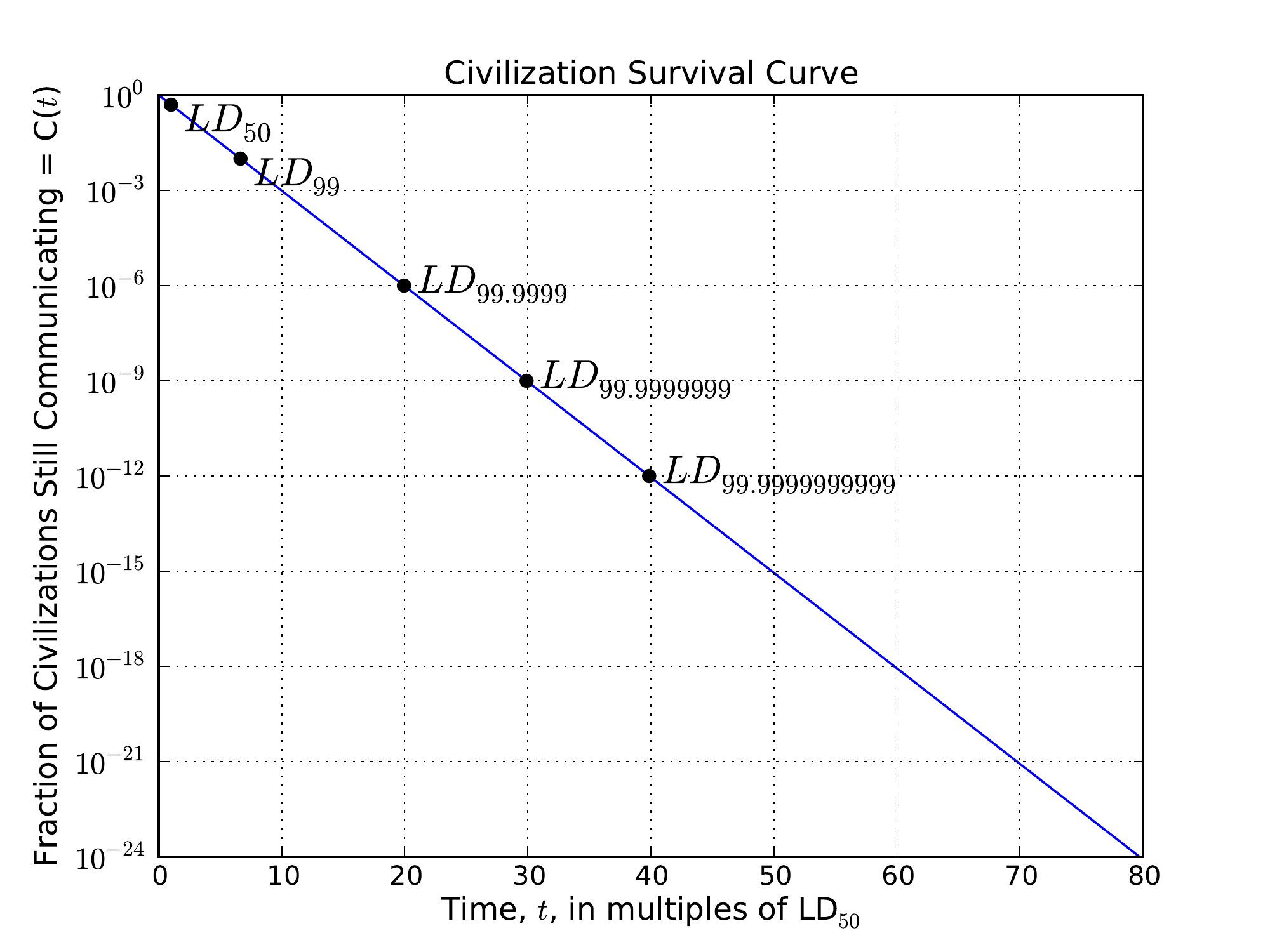}}
    \mycaption{Survival times in a cohort of civilizations, all created at $t=0$.}{\textsl{Left:} Over time, the percentage of silent civilizations, $100(1-C(t))$, logarithmically approaches 100\%. For any $E$ and $P$, $LD_{X\%}=\frac{ln(1-X\#)}{ln(1-0.50)}LD_{50}$, where $X\%$ is a percentage and $X\#$ is the equivalent probability. \textsl{Right:} This panel modifies the left panel's axes. First, the time axis is expanded compared to the left.  Second, the vertical axis has been inverted to show survival, $C(t)$, over time. The $LD_{50}$ and $LD_{99}$ points carry over from the left panel.  Remarkably, the time required to reach infinitesimal survival rates, e.g., $10^{-24}$, is less than two orders of magnitude larger than the median civilizational survival time, $LD_{50}$.}
    \bigskip
    \hrule
    \label{fig:LDLines}
    \end{figure}

\noindent Increasing the certainty of civilizational death increases the lethal duration exponentially, as Figure~\ref{fig:LDLines} shows.  Thus, for any $E$ and $P$, $LD_{95}\approx (4.3\ LD_{50})$, $LD_{99.9999}\approx (20\ LD_{50})$, and $LD_{100[1-C(y)]}\approx (80\ LD_{50})$ where $C(y) = 10^{-24}$.

\begin{figure}[!ht] %%%%%%%%% Figure 2
    \ \vskip-2ex
    \centerline{\includegraphics[scale=1]{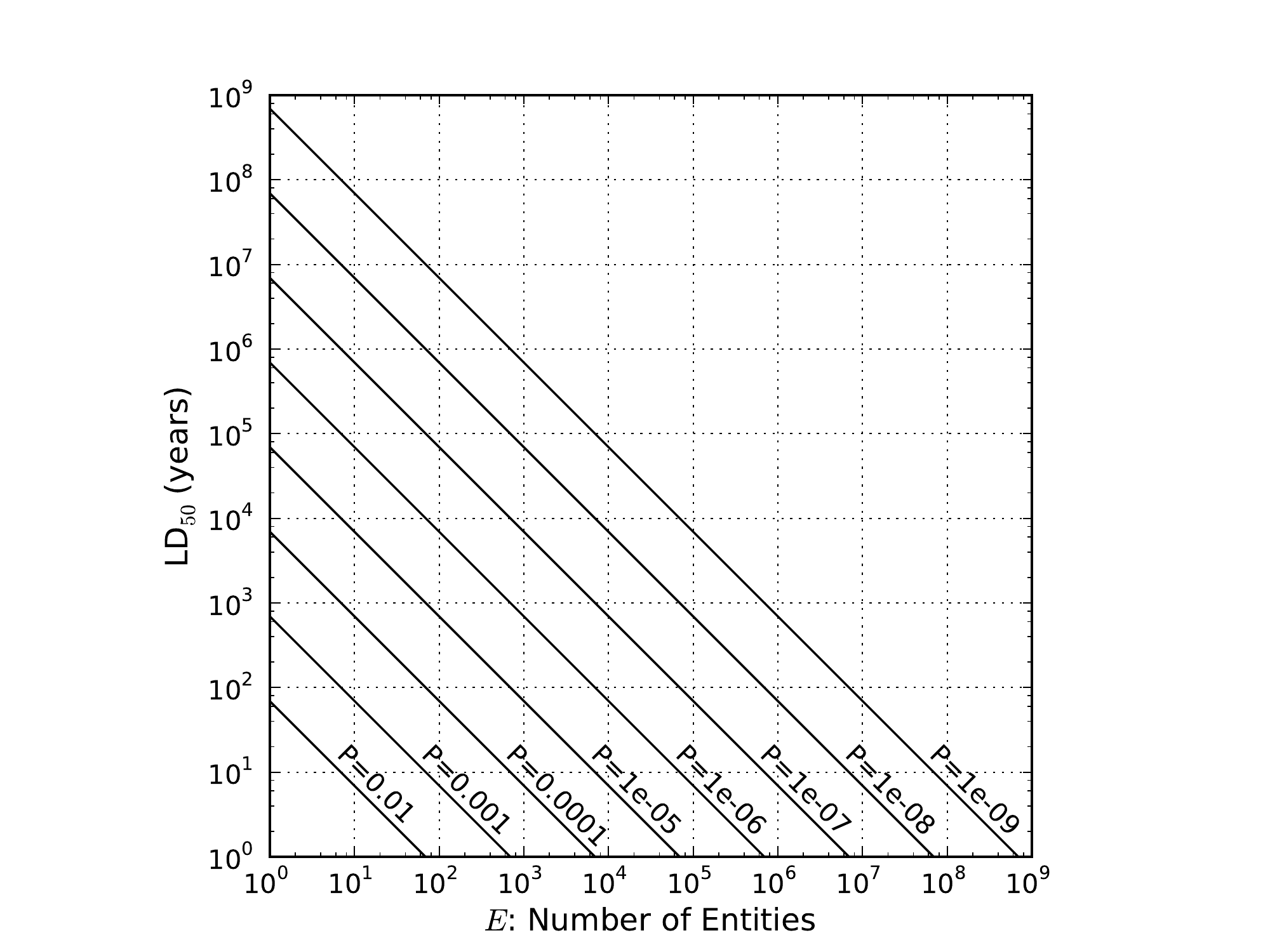}}
    \mycaption{Technology diffusion ($E$) and psychosociology ($P$) determine civilizational lifespan ($LD_{50}$).}{$E$ is the number of entities who control a means to end civilization. $P$ is the probability per annum per entity that the entity will trigger its civilization-ending means. Given a constant $E$ and $P$, $LD_{50}$ is the median number of years before civilization is expected to end.  $E$ and $LD_{50}$ have an inverse linear relationship for any $P$.}
    \bigskip\hrule
    \label{fig:EPLines}
    \end{figure}

\def\LDfiftyCoeff{0.7}
Figure~\ref{fig:EPLines} plots the relationship between $E$ and $LD_{50}$ for several $P$, and illustrates the approximation $LD_{50} \approx {\LDfiftyCoeff}/({E\times P})$, derived in Equation~\ref{eqn:LDFiftyCoeff} of the Mathematical Appendix.

\newcommand*\mean[1]{\overline{#1}}
\def\meanC{\mean{C(y)}_W}

\begin{figure}[!ht] %%%%%%%% Figure 3
    \ \vskip-8ex
    \centerline{\includegraphics{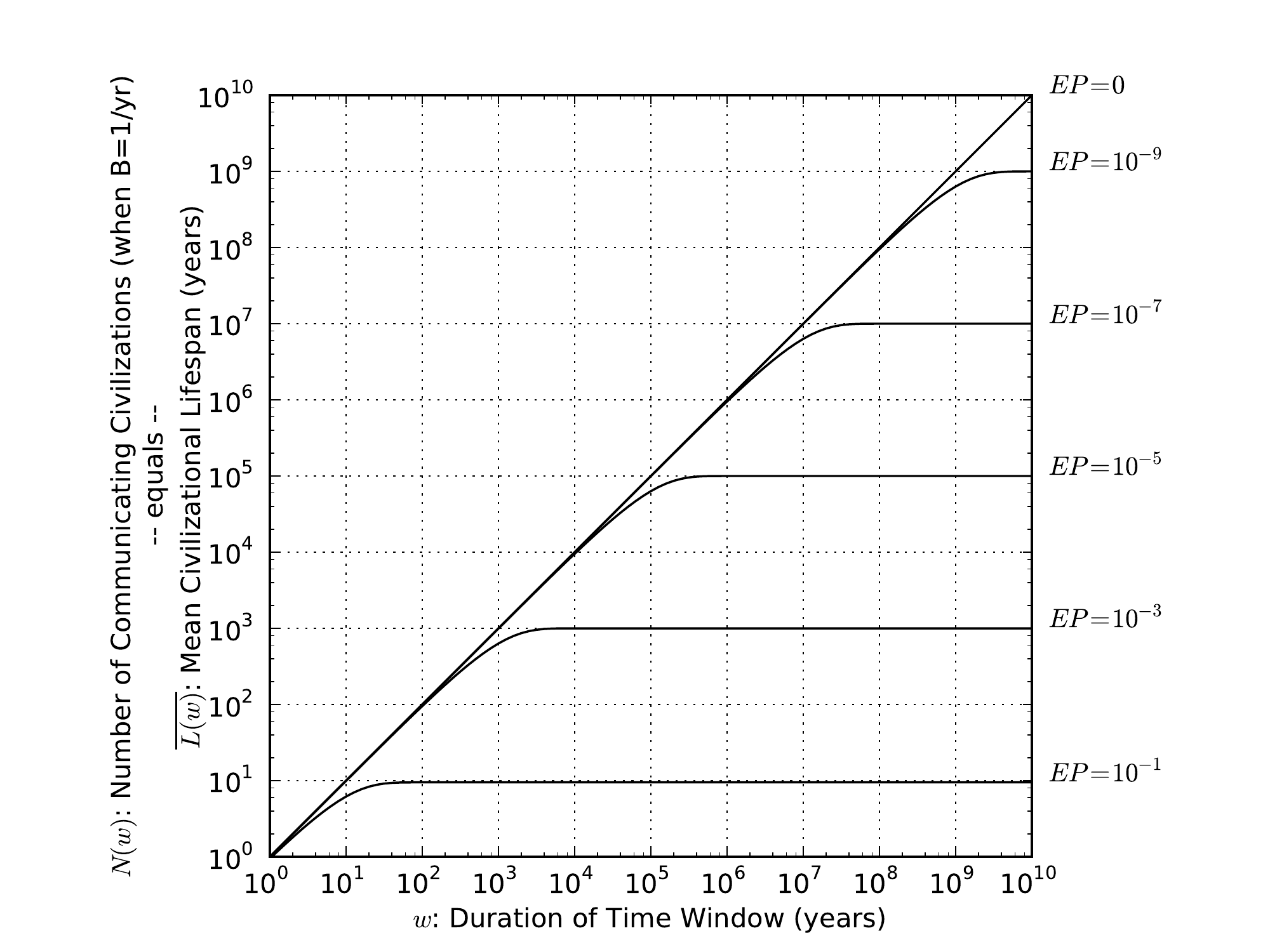}}
    \mycaption{Civilizations and time.}{For six different values of $E\times P$, the plot shows two equivalent quantities \revised{for time windows of various durations $w$}: (a)~$\Lww$ = mean lifetime of communicating civilizations over time, and (b)~$N(w)$ =  number of communicating civilizations over time when $B=1$. For both quantities, a constant $B$ is assumed. Zero civilizations exist at time $w=0$.  Equation~\ref{eqn:maxLwwLimit} mandates $\Lww<1/(EP)$ for all $w$. Per Equation~\ref{eqn:Nt-approx}, $\Lww$ grows substantially until a near-steady-state is reached at about $w=10/(EP)$ years.  An arbitrary-precision software package (Johansson et al.\ 2014) used Equation~\ref{eqn:Nt-Exact} to calculate $N(w)$ and $\Lww$.}
    \bigskip\hrule
    \label{fig:civcounts}
    \end{figure}

To calculate the mean lifespan, it is more intuitive to first calculate the number of communicating civilizations, $N(w)$, that exist at the end of a time window extending from year $y=0$ to $y=w$.  Assuming that zero civilizations existed at $y=0$, and that communicating civilizations were born at a constant rate of $B$ per year throughout the time window, the Mathematical Appendix shows:
\begin{align*}
N(w)&=B \int_0^{w} C(y)\ dy\tag{is \ref{eqn:integNt}}\\
\\
&=B \frac{S^w-1}{ln\ S}\ \textrm{where}\ S=(1-P)^E\tag{is \ref{eqn:Nt-Exact}}\\
\\
&\approx \frac{B}{EP}\forWsmallP\tag{is \ref{eqn:Nt-approxB}}
\end{align*}
\noindent Figure~\ref{fig:civcounts} plots the exact form of $N(w)$ from Equation~\ref{eqn:Nt-Exact}, for multiple $w$ and $EP$ when $B=1$. It shows with reasonable precision that $N(w)\leq B/(EP)$ for any $w$.

The parameter $L$ in the Drake equation is reformulated herein to $\Lww$, the mean lifespan for civilizations born during a time window of duration $w$.  This transforms the Drake equation to:
\vskip -3ex
\begin{align*}
N(w) = B\ \Lww\tag{is \ref{eqn:NBL}}
\end{align*}
Thus, $\Lww=N(w)$ when $B=1$, and so Figure~\ref{fig:civcounts} is also a plot of $\Lww$.  

Per Figure~\ref{fig:civcounts}, $\Lww$ increases with $w$.  However, its maximum value, at any time, is constrained.  \revised{Assuming all civilizations have identical $E$ and identical $P$:}

\vskip -3ex
\begin{align*}
\Lww_{max} < \frac{1}{EP}\ \ \ \ [\textrm{for all}\ w]\tag{is \ref{eqn:maxLwwLimit}}
\end{align*}

Combining these two formulae and defining $N$ as ``$N(w)$ for all $w$'' yields the Drake equation as an inequality:
\begin{equation}
N < \frac{B}{EP}
\label{eqn:mydrake}
\end{equation}
\noindent\revised{or, hewing to its classical form \cyte{drake1961}:}
\begin{equation}
N < \frac{R_* f_p n_e f_l f_i f_c}{EP}
\end{equation}
\noindent Because the model addresses only endogenous involuntary silencings, adding consideration of other causes for silencings would merely reinforce this inequality.

To produce near-term risk estimates for Earth, a PubMed search informed the value of $E$, as follows.   With the assumption of a civilization-ending technology based on some yet-to-be-described genetic technique, the number of people authoring scientific articles indexed under ``genetic techniques'' (one of PubMed's $\approx$27,000 standard index terms) can be used to estimate the number of people capable of exploiting such a technique, thereby serving as a proxy for $E$.  Thus, the PubMed search

\smallskip
\centerline{\texttt{genetic techniques[mh] AND "2008/01/01"[PDAT]:"2015/12/31"[PDAT]}}
\smallskip

\noindent performed on August 10, 2017, yielded 594,458 publications in the most recent eight-year span of complete bibliographic coverage. After eliminating non-scientific publications (of type letter, comment, news, interview, etc.) 585,004 remained, which carried 1,555,661 unique author names.  Of these authors, approximately 179,765 appeared on five or more publications.  This number is a maximum because some authors publish under more than one name.

\revised{Models employing non-constant $E$ and $P$ are possible.  The simplest posits that $E$ grows as population might: a fixed percent per year.  If, over $y$ years, $E$ grows this way from some initial value $E_0$, with the growth continuously compounded, then:}

\begin{equation} 
E_y = {E_0}\thinspace e^{ry}
\label{eqn:defEgrowth}
\end{equation}

\noindent\revised{where $r$ is the growth factor (e.g.\ 0.02 for 2\%\ annual growth) and $e=2.71828...$.  Unfortunately, the unbounded exponential term renders this ``growth model'' nonsensical for even moderately large $y$. Still, some insights can emerge for short time horizons, as detailed in Figure~\ref{fig:growthmodel}, which is based on Equation~\ref{eqn:yOneyTwo} in the Appendix. Unsurprisingly, a growing $E$ yields an $LD_{50}$ significantly smaller than is calculated from a constant $E$.}

    \begin{figure}[!ht] %%%%%%%% Figure 4
    \ \vskip-8ex
    \centerline{\includegraphics{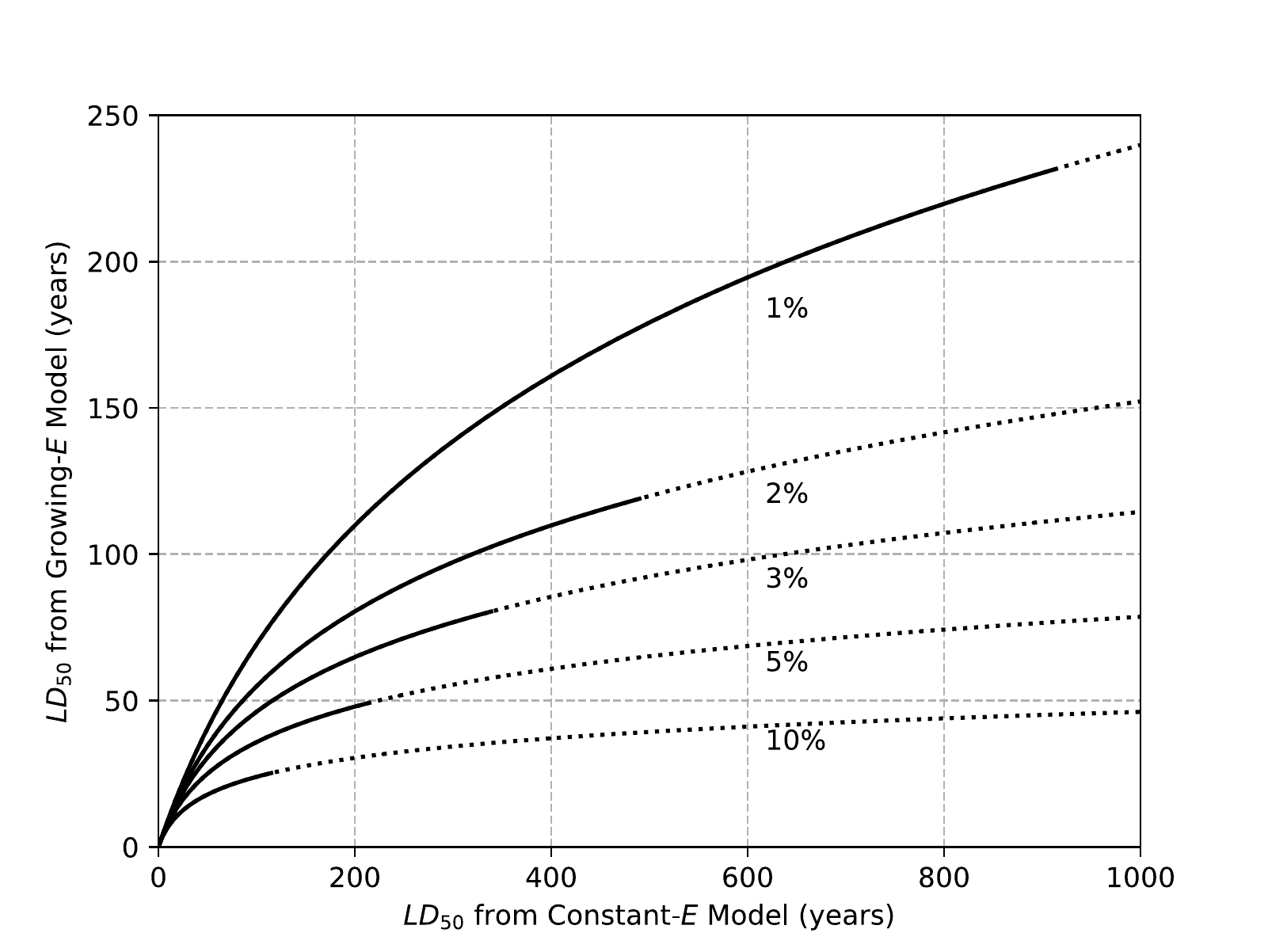}}
    \mycaption{Drop in $LD_{50}$ when $E$ grows.}{\revised{The horizontal axis corresponds to $LD_{50}$ values calculated from Equation~\ref{eqn:defLDFifty} and a constant $E$ and $P$.  If, however, $E$ is not constant, and instead grows at a fixed percentage annually (five growth rates are shown), then $LD_{50}$ shrinks to the corresponding value on the vertical axis, according to Equation~\ref{eqn:LDFiftyEq}.  So, for example, an $LD_{50}$ of 600 years derived from Figure~\ref{fig:EPLines} would be revised to approximately 190 years if $E$ grew by 1\%\ annually. To signal wariness about exponential explosion, each solid line changes to a dotted line when the number of entities has increased a million-fold (i.e., $E_y/E_0 \ge10^6)$.}}
    \bigskip\hrule
    \label{fig:growthmodel}
    \end{figure}

%%%%%%%%%%%%%%%%%%%%%%%%%%%%%%%%%%%%%%%%%%%%%%%%%%%%%%%%%%

\section{Discussion of Model}

\revised{Unless explicitly noted, all discussion refers to the baseline model in which $E$ is constant.}

Equation~1 provides the probability, $C(y)$, that a civilization survives endogenous involuntary silencing threats until some $L=y$.  The lethal durations, $LD_{50}$ et al, are probabilistic statements of this $L$.  Because the model terminates upon the first use of a civilization-ending technology, more complicated models, such as the Poisson distribution, are not required.

Thus, the model is simple, but not unreasonably so.  With only two parameters, however, it is important to understand their inherent assumptions.

\subsection*{Model Discussion: $E$ and $P$ (and $B$)}

In broad terms, $E$ characterizes a CE technology and its availability, while $P$ characterizes the psychology and sociology of the entities who possess the technology.  Although loss of interstellar communicativeness is equated to the end of civilization, other endpoints (e.g., complete extinction) could be substituted.  The only criteria are consistency of the endpoint, independence of the entities, and termination of the model upon first triggering.

Numerous subtleties attend the definitions of $E$ and $P$.

First, $E$ applies to any CE technology, be it nuclear, nano-, bio-, or another.  The CE technology never fails to end civilization once triggered.  The effects of ``near miss'' extinction events on population and psychology are ignored.

Second, $E$ includes only entities that possess (or can acquire) the ``full stack'' of CE technology.  That is, they must have the capability to make or otherwise obtain the weapon, and to deliver it in quantities that render the civilization uncommunicative.  So,  for example, even though designs for nuclear weapons are comparatively well known \cyte{phillips1978}, $E$ for Earth remains only $\approx2$ (representing the leaders of the United States and Russia).\footnote{$E$ would be slightly higher if additional other leaders could end civilization via climate change.}  The self-propagating nature of biological weapons would simplify, but not eliminate, the delivery challenge. 

Third, to the extent that machine intelligences possess CE technology, they could also be counted in $E$. (Exemplar: ``SkyNet'' from the \textsl{Terminator} movies.)

Fourth, $E$ reflects a balance between offensive and defensive technologies.  Thus, developing and readying defensive technology offers a straightforward, albeit challenging, path to markedly decrease $E$.

Fifth, $P$ is the sum across all reasons, intended or not, that an entity might trigger the CE technology.\ Most are psychosocial, e.g., greed, hate, stupidity, folly, gullibility, power-lust, mental illness, ineptitude, non-fail-safe design, etc. \textsl{The Bulletin of the Atomic Scientists}' ``doomsday clock'' \cyte{bulletin2002} has similarities to $P$.

Sixth, the model assumes constant $E$ and $P$ throughout the time window of interest.  This is unlikely to occur in a real civilization, given the dynamics of offensive/de\-fensive technologies, population, sociopolitical stability, and technology diffusion.  Simple model extensions would have $E$ and $P$ vary over time, or sum across subpopulations of entities each with their own $E_i$ and $P_i$, or sum across multiple CE technologies each with their own $E_j$ and $P_j$.\footnote{The model would become complex to the extent that interaction terms would be needed to model a single entity having access to multiple CE technologies.  However, modeling the technologies separately and then choosing the most pessimistic outcome would likely suffice.} 

\revised{Unlike Cooper \cyte{cooper2013} population growth -- and concomitant growth in $E$ -- is omitted from the baseline model because all realistic non-zero growth rates become nonsensical when compounded (exponentiated) over eons. Over short time frames, the effect of a growing $E$ can be reasonably equated to a speed-up in time.  For example, when $E$ is constant and the model reaches some state at year $y$, a situation in which $E$ is growing by 2\%\ annually will attain the same state significantly earlier, at year $50\thinspace ln\thinspace (1+y/50)$ according to Equation~\ref{eqn:LDFiftyEq}.}

The model's flexibility could be improved -- at the cost of great mathematical complexity -- by assigning probability distributions to $E$ and $P$ and convolving them. However, models that assume a distribution around some mean value for $P$ (denoted $P_{mean}$)
%\footnote{\OurModel\ may be said to use a Kronecker delta distribution for $E$ and $P$.} 
will yield lower values for $C(y)$ and $LD_{50}$ than the present model, because of the positive exponent in the definition of $C(y)$.  Thus, this model's dispiritingly low values for $LD_{50}$ nevertheless represent a civilization's best-case outcome for a given $P_{mean}$.

This is most obviously appreciated in the edge case where a single entity has its $P=1$, for example, an entity who acquires the skills of a CE technology specifically to end civilization.  As soon as a single qualified entity has $P=1$, then the overall civilizational $P$ is also 1, and $LD_{50}$ (in fact, all $LD_x$) is zero.

Civilizations spanning multiple planets should be treated as multiple civilizations, each modeled separately with their own $E$ and $P$.  Modeling them as a single civilization assumes all the planets' civilizations die from one attack -- an unnecessarily stringent requirement.  Of course, $P$ might change on planets that see a sister planet destroy itself.

Although colonization would imply a non-constant $B$, the model would still apply so long as $B$ is less than some constant $B_{max}$.  Using $B_{max}$ in the model would provide an upper bound for $N(w)$.  Geometrically increasing $B$ would require re-working the model, but the barrenness of the galaxy mitigates this possibility: Tipler \cyte{tipler1980} and others \cyte{webb2015} \cyte{jones1981} \cyte{armstrong2013} note that a single civilization colonizing at even moderate rates of geometric increase would fill the galaxy in only a few million years, and we do not observe a full galaxy.

Furthermore, assuming that the technology of interstellar colonization is far more daunting than biotechnology, and that the self-preservation drives of individual intelligences far exceed any elective desire to migrate off-planet, it is reasonable to expect that, as a rule, civilizations will develop and use sophisticated biotechnology before dispersing themselves on other planets \cyte{cooper2013}. Thus, the experience of 20th century Earth is likely typical, i.e., the progress of medicine and public health in the era antedating genetic biotechnology creates a population explosion, so that civilization consists of a large, dense, mobile population on a single home world at the time that potentially CE biotechnology is developed.  Because such ecological conditions are conducive to the spread of communicable agents, it is reasonable to hypothesize that all planetary civilizations will face existential threats from contagious micro-organisms -- whether engineered or not -- before they become vigorous interstellar colonizers \cyte{cooper2013}.

\OurModel\ could also apply to civilizations based on networked machine intelligences when epidemic malware is a possibility. Because diversity among evolution-produced organisms would likely be higher than among designed software, building CE technology against machine intelligences could be comparatively easy.

\subsection*{Model Discussion: Stability}

It may be argued that a potential CE technology cannot exist for long time spans without a defensive technology being developed, i.e., that $E$ cannot exceed zero for thousands, millions, or billions of years.

Several considerations weaken this proposition, especially as relates to biotechnology.  These considerations are illustrative, and necessarily speculative.  Future biotechnological progress will elucidate the extent to which they hold.

First, reliance on a single CE technology is not required.  Instead, multiple CE technologies may exist serially, each enabling a multitude of different attacks, with each attack requiring a different defense.  This is akin to the inventory of ``zero day exploits'' that present-day entities accumulate to penetrate computer systems. 

Second, a long period of $E>0$ can be viewed as the concatenation of shorter time periods having $E_i>0$, where each $E_i$ derives from a separate CE attack possibility that is eventually countered by a defense tailored to that attack.  For example, if the frailties of life allow for a million different attacks,\footnote{Even simple viruses have profound combinatorial reserve. Influenza~A, for example, with its genome of $\approx$14,000 nucleotides, has $\approx$880 million combinatorial two-nucleotide variants and $\approx$12 trillion three-nucleotide variants \cyte{perelson2012}. Though only a sliver of these would yield functionally and/or immunologically distinct viruses, the numerator explodes exponentially. It is a tall order to devise anti-influenza~A technologies that are 100\%\ effective against all possible variants.} and it takes one year to tailor a defensive technology for each, then $E>0$ for $w=10^6$ years. If no periods of $E=0$ were interspersed between the $E_i>0$ periods, then the time window $w$ would equal elapsed time in the universe. In scenarios having interspersed $E_i=0$ periods, elapsed time would exceed window duration.

Third, mere development of defensive technology is not sufficient.  The technology must be fully fielded.  That is, unless widespread pre-exposure vaccination is possible, an attack must be detected, the agent(s) characterized, and the remedy developed, tested, manufactured (perhaps in billions of doses), distributed, and administered -- all of which must succeed before the attack can take root in the population.  This is a formidable challenge requiring multiple sub-technologies in the near term, or a single future technology that is currently indistinguishable from magic.

Fourth, defensive technology may be impossible on first principles.  For example, every known life form adapts its gene expression to its environment.  An offensive technology whose only defense necessitated extinguishing this genetic responsiveness would seem unobtainable.

Fifth, mere possession of defensive technology is not sufficient -- timely and correct decisions to activate defenses on a civilizational scale must also occur. Thus, a civilization's decision-making process, be it political, machine-based, or other, is also a target for CE technologies. This means $E$ has a small psychosociological component.\footnote{Alternatively, the model could divide $P$ into $P_{offense}$ and $P_{defense}$.} Decentralized decision-making, such that every individual intelligence possessed counter-CE technology and independently decided when and if to self-medicate, would require a level of trust in the population that no government on earth has so far developed.

Sixth, generalizing the above scenario, CE technologies need not be highly lethal.  To sustain itself, a densely populated world may rely on critical infrastructure and/or heavily optimized industrial processes.  Direct or indirect disruption of these essential functions could cause sufficient social chaos to render a civilization uncommunicative.

Finally, if $EP$ is large throughout the universe, then the model does not have to apply for millions or billions of years.  For example, if $E=10^3$ and $P=10^{-3}$ then $LD_{50}\approx 0.7$ years and the probability of surviving to 25 years is $<10^{-9}$.

%%%%%%%%%%%%%%%%%%%%%%%%%%%%%%%%%%%%%%%%%%%%%%%%%%%%%%%%%%

\section{Discussion of Results}
\subsection*{Results Discussion: Earth}

From Equation~\ref{eqn:LDFiftyCoeff}, achieving $LD_{50}\geq 1000$ years requires $EP\leq 7\times 10^{-4}$.  Thus, with $E=2$ today, $P\leq .00035$ is required.

Given the pace of biotechnology's progress, plus the irresistible pressure to continue that progress for universally-desired medical purposes, plus the dual-use potential of the technology, plus its potential worldwide reach, 
many humans could soon have the capacity to end Earth's technical civilization, driving $E\gg 2$.  In a recent eight-year span, more than 1.5 million people participated in the ``genetic techniques'' enterprise at a level sufficient to warrant authorship on a scientific article. Almost 180,000 of them authored five or more such articles. The number actually engineering artificial organisms today is certainly far smaller, but clearly a large reservoir of hands-on molecular genetics competence already exists on Earth.

Although $LD_{50}$ has been our focus, planning with lower thresholds \cyte{suskind2006}, e.g., $LD_{05}$ ($\approx LD_{50}/13.5$) or $LD_{01}$ ($\approx LD_{50}/70$), would mitigate unanticipated rapid rises in $E$ or $P${}. For example, comparing a CE technology's $LD_{01}$ to the anticipated time needed to develop defensive counter-technology might drive policy makers to speed such development.  

Given the PubMed authorship numbers, a few new biotechnological innovations could reasonably and quickly raise $E$ to $10^4$.  If so, and $P=10^{-7}$, then $LD_{01}\approx10$ years.  If $E$ became larger, $LD_{01}$ would become smaller.  The short $LD_{01}$ time span is concerning, given today's comparatively slow pace of anti\-microbial innovation (the common cold and many other infections remain incurable and without vaccinations), and strongly argues that defensive technology development must be expanded and must occur simultaneously with any therapeutic (offensive) development.

An especially concerning scenario arises if, someday, hospitals employ people who routinely write patient-specific molecular-genetic programs and package them into replicating viruses that are therapeutically administered to patients, especially cancer patients.  If the world attained the European Union's per capita hospital density,\footnote{In 2004, 15,000 hospitals \cyte{hope2009} were serving 500 million people \cyte{oecd2009}. Likely, few emerging health systems will follow an American model.} this could mean two hundred thousand hospitals employing perhaps 1 million people who might genetically engineer viruses every workday.  Should techniques emerge for a highly communicable therapeutic virus -- against which vaccination would be refused, as that would preclude future cancer therapy -- and $E$ reached $10^6$, then attaining an $LD_{01}$ of just 10 years would require $P<10^{-9}$, perhaps an impossibility, given human nature.

%%%%%%%%%%%%%%%%%%%%%%%%%%%%%%%%%%%%%%%%%%%%%%%%%%%%%%%%%%

\subsection*{Results Discussion: Drake Equation}

By simulating an ensemble of civilizations, the present model challenges Burchell's assertion \cyte{burchell_2006} that $L$ in the Drake equation is ``not truly estimable [estimatable] without observation of a set of societies.''  Although estimating $P$ based on first principles cannot be done for extraterrestrial civilizations, estimating $E$ and the product $EP$ may be tractable\revised{ within the assumptions of the model, as follows}.

Lower-bound estimates for $E$ would derive from deep understanding of the genetic mechanisms of life -- all possible mechanisms, not just DNA/RNA -- and from the possibilities of biotechnology as applied to those mechanisms. \revised{Thus, estimates of $E$ would derive from understanding the gamut of intelligence-compatible biologies, an understanding that smart human biochemists could perhaps achieve {\sl ex nihilo}, without interstellar travel or communication.} Machine intelligences would have analogous considerations.  The existence of other CE technologies might increase $E$ further. 

Because of Equation~\ref{eqn:mydrake}, $EP$ can be constrained by searching for extraterrestrial intelligence (SETI). With $B$ increasingly well understood, constraining $N$ in Equation~\ref{eqn:mydrake} constrains $EP$.  Thus, if SETI efforts someday yielded a conclusion such as ``We estimate that no more than $N_x$ communicating civilizations exist,'' then $EP<B/N_x$. 

If both $EP$ and $E$ can be estimated, then the value of $P$ is constrained. \revised{It is interesting to note that, given its dependence on psychological factors, possessing a constraint or estimate of $P$ would be a first step toward a quantitative epidemiology of alien psychologies.}

The model applies so long as opportunities to deploy civilization-ending means predate the ability to counter all such attacks (and accidents).  That is, whenever $E>0$, Equation~3 produces a finite value for $LD_{50}$ and civilization is at risk, assuming $P>0$. Whether any measures could achieve $P=0$, short of pervasive and perfect surveillance of entities, is unknown.

The model's low values for lifespan, $\Lww$, have implications for SETI strategy.  If geometrically increasing interstellar colonization circumvents short civilizational lifespan, then, \revised{all other factors being equal, communicating civilizations would be longest-lived where such colonization is easiest, e.g.\ where the time and/or energy required to move between habitable planets is smallest.
This consideration adds to existing reasons why SETI might target zones of densely collected habitable planets \cyte{turnbull2003}.}

%%%%%%%%%%%%%%%%%%%%%%%%%%%%%%%%%%%%%%%%%%%%%%%%%%%%%%%%%%

\subsection*{Results Discussion: the Fermi Paradox and the Great Filter}

To date, in a visible universe of $\approx 10^{24}$ stars and their planets, only Earth shows evidence of intelligent life.  This apparent paradox, noted by Enrico Fermi and others \cyte{webb2015}, could be explained by a ``Great Filter'' that all but prevents communicating civilizations from forming or surviving \cyte{hanson2009}. The Great Filter may be technological in origin if ``(a) virtually all sufficiently advanced civilizations eventually discover it and (b) its discovery leads almost universally to existential disaster''  \cyte{bostrom2008}.

\iffalse
    Biotechnology is inescapably silencing because it is inherently dual-use: 
 \fi

Most remarkably, the present model supplies the quantitative 24 orders-of-magnitude winnowing required of a Great Filter, reducing it to a two-orders-of-magnitude multiplication.  For example, if $E=10^6$ and (optimistically) $P=10^{-9}$, then $LD_{50}\approx700$ years, and $LD_{100[1-C(y)]}\approx (80\ LD_{50})\approx 56,000$ years when $C(y)=10^{-24}$. That is, for this $E$ and $P$, we expect only one civilization in $10^{24}$ to still be communicating after 56,000 years, and even a galactically-short 100,000-year lifespan is effectively impossible because only one in $10^{42}$ civilizations remains communicative.

\def\mymathhyphen{{\hbox{-}}}
\def\subthree{elective}
\def\subfour{{non\mymathhyphen chaotic}}
\def\subfour{{other}}

Overall, therefore, \we\ would advise advanced technical civilizations to optimize not on megascale computation \cyte{sandberg2017} nor engineering \cyte{dyson1960} nor energetics  \cyte{kardashev1964}, but on defense from individually-possessable self-replicating existential threats, such as microbes or nanomachines.

%%%%%%%%%%%%%%%%%%%%%%%%%%%%%%%%%%%%%%%%%%%%%%%%%%%%%%%%%%

\bigskip
\textbf{Acknowledgments: } I am grateful for the support and wise counsel of Jennifer Esposito, Mike Morton, Barry Hayes, and, of course, Tanya Roth.  However, all errors are the author's responsibility. \revised{My thanks go to the anonymous reviewer for spurring development of Figure~\ref{fig:growthmodel}.} 

%%%%%%%%%%%%%%%%%%%%%%%%%%%%%%%%%%%%%%%%%%%%%%%%%%%%%%%%%%
\vfill\break
\section*{Mathematical Appendix}
\newcounter{mathsectionCounter}
\def\mathsubsection#1{\ifnum\value{mathsectionCounter}>0\vfill\break\fi\stepcounter{mathsectionCounter}\subsection*{\hrule\smallskip\noindent Math \arabic{mathsectionCounter}: #1\smallskip\hrule}}

% Put an "A" in front of equation numbers in the appendix.
% From: https://tex.stackexchange.com/questions/118606/numbering-tables-a1-a2-etc-in-latex
\setcounter{equation}{0}
\renewcommand{\theequation}{A\arabic{equation}}

%=================================================
\mathsubsection{$ln(1-P)=-P$ as $P\rightarrow 0$}

To solve
$$f(P)=\frac{ln(1-P)}{P}$$

\noindent at $P=0$, we observe that $f(0)$ evaluates to 0/0, making the expression indeterminate.  However, it also means L'H\^opital's rule applies in the second step below:
\begin{align*}
\lim\limits_{P\rightarrow 0} f(P)&=\lim\limits_{P\rightarrow 0} \frac{
    ln(1-P)
    }{
    P
    }\\
\ \\
&=\frac{
    \lim\limits_{P\rightarrow 0} \frac{d[\ ln(1-P)\ ]}{dP}
    }{
    \lim\limits_{P\rightarrow 0} \frac{d[\ P\ ]}{dP}
    }\\
\ \\
&=\frac{
    \lim\limits_{P\rightarrow 0} (\frac{-1}{1-P})
    }{
    \lim\limits_{P\rightarrow 0} (1)
    }\\
\ \\
&=\frac{-1}{1}\\
\end{align*}
Hence:
$$\lim\limits_{P\rightarrow 0} \frac{ln(1-P)}{P}=-1$$

\medskip
\noindent So, when $P\rightarrow 0$ we can use:
\begin{equation}
ln(1-P)=-P
\label{eqn:PlnP}
\end{equation}

\noindent\revised{For our purposes this approximation is excellent, viz.\ $ln(1-0.1)=-0.105$ and $ln(1-0.001)=-0.0010005$.}

%=================================================
\mathsubsection{$LD_{50}$ as $P\rightarrow 0$}

We start with Equation~\ref{eqn:defLDFifty} defining $LD_{50}$, then simultaneously take the limit and substitute Equation~\ref{eqn:PlnP} into it:

\begin{align}
LD_{50}& = \frac{ln(1-.50)}{E \times ln\ (1-P)}\nonumber\\
\nonumber\\
\lim\limits_{P\rightarrow 0} LD_{50}& = \frac{ln(1-.50)}{E\times (-P)}\nonumber\\
\nonumber\\
& = \frac{ln\ 2}{E\times P}\label{eqn:LDFiftyEP}\\
\nonumber\\
& \approx \frac{\LDfiftyCoeff}{E\times P}\label{eqn:LDFiftyCoeff}
\end{align}

Thinking solely in terms of exponents:
$$LD_{50} \approx .7\times 10^{-(log_{10} E\ +\ log_{10} P)}$$

%=================================================
\mathsubsection{$N(w)$ -- Exact}

Recall from Equation~\ref{eqn:defCy} that $C(y)$ is the fraction of civilizations still communicating $y$ years after their birth.  Here, however, the notion of time changes a bit.

First, define:
\begin{align*}
B(y)& = \textrm{number of new civilizations \underline{b}orn in year } y\\
N(y)& = \textrm{number of communicating civilizations existing in year } y\\
\end{align*}

\vskip -3ex
Next, assume we are interested in a \underbar{w}indow of time in the galaxy's history running from year 0 to year $w$, where no civilizations were present at $y=0$.  We want to know the number of communicating civilizations that exist at the end of the window, i.e. at time $w$.  

To be considered alive at year $w$, any civilization born in some year $y$ will have to communicate for $w-y$ more years.  Thus:

$$N(w)= B(0)\ C(w) + B(1)\ C(w-1) + B(2)\ C(w-2) + ...+ B(w)\ C(0)$$

Assuming $B(y)$ is a constant (having units: civ year$^{-1}$):

\begin{equation}
N(w) = B \sum_{y=0}^{w} C(y)
\label{eqn:sumNt}
\end{equation}

We can replace summation with integration:

\begin{equation}
N(w)=B \int_0^{w} C(y)\ dy
\label{eqn:integNt}
\end{equation}

To solve for $N(w)$, \revised{assuming all civilizations have the same $E$ and $P$}, we define:

\begin{equation}
S=(1-P)^E
\label{eqn:S-definition}
\end{equation}

Substituting the above into Equation~\ref{eqn:defCy} yields:

\begin{equation}
C(y)=S^y
\label{eqn:Cy-for-S}
\end{equation}

Then substituting Equation~\ref{eqn:Cy-for-S} into Equation~\ref{eqn:integNt}:

\begin{align*}
N(w)&=B\int_0^{w} {S}^y\ dy\\
&=B\ \frac{S^y}{ln\ S}\bigg\rvert_0^w\\
&=B\bigg(\frac{S^w}{ln\ S}-\frac{S^0}{ln\ S}\bigg)\\
\end{align*}
This yields the exact form of $N(w)$:
\begin{equation}
N(w)=B\ \frac{S^w-1}{ln\ S}\quad \textrm{where}\quad S=(1-P)^E
\label{eqn:Nt-Exact}
\end{equation}

%=================================================
\mathsubsection{$N(w)$ -- As $P\rightarrow 0$ and $w\rightarrow \infty$}

In many scenarios for $N(w)$, $P\rightarrow 0$ and/or $w\rightarrow\infty$. We here derive an approximation for such conditions.

First, expand the exact definition of $N(w)$ in Equation~\ref{eqn:Nt-Exact}:

\begin{align*}
N(w)&=B\ \frac{((1-P)^E)^w - 1}{ln\ ((1-P)^E)}\\
\\
&=B\ \frac{(1-P)^{(Ew)} - 1}{E\ ln\ (1-P)}\\
\end{align*}

Now substitute with the results of Equation~\ref{eqn:PlnP}, namely $ln(1-P)=-P$ when $P$ is small and, consequently, $(1-P)=e^{-P}$:

\begin{align*}
\lim\limits_{P\rightarrow 0} N(w)&= B\ \frac{e^{-PEw} - 1}{E\ (-P)}\\
\\
&= B\ \frac{1-e^{-PEw}}{EP}\\
\end{align*}

\noindent As $w$ becomes large, $e^{-PEw}\rightarrow0$.  Thus:
% \nonumber from: 
%https://tex.stackexchange.com/questions/17528/show-equation-number-only-once-in-align-environment
\begin{equation}
\lim\limits_{\substack{P\rightarrow 0\\ w\rightarrow\infty}}N(w)=\frac{B}{EP}
\label{eqn:Nt-approx}
\end{equation}

Using Figure~\ref{fig:civcounts}, which was calculated using the exact form of $N(w)$ in Equation~\ref{eqn:Nt-Exact}, we observe the approximate value-range of $w$ for which the limit of Equation~\ref{eqn:Nt-approx} holds:

\begin{equation}
N(w)\approx\frac{B}{EP}\forWsmallP
\label{eqn:Nt-approxB}
\end{equation}

%=================================================
\mathsubsection{$\Lww$}

As many others have noted, the Drake equation can be reduced to a two-parameter form:
\begin{align}
N = B \times L
\label{eqn:DrakeTwo}
\end{align}

\noindent where $N$ is the number of communicating civilizations, $B$ is the birth rate of communicating civilizations, and $L$ is the mean lifetime of all birthed civilizations.

Applying this to our approach of examining time windows having constant $B$, we can rewrite Equation~\ref{eqn:DrakeTwo} as:
\begin{align}
N(w) = B \times \Lww
\label{eqn:NBL}
\end{align}

\noindent where $\Lww$ is the mean lifetime of a civilization born during the time window that extends from 0 to $w$.

Rearranging Equation~\ref{eqn:NBL} and then substituting from Equation~\ref{eqn:Nt-Exact} yields:
\begin{align}
\Lww &= \frac{1}{B}\ N(w)\nonumber\\
\nonumber\\
&= \frac{1}{B}\ B\ \frac{S^w-1}{ln\ S}\nonumber\\
\nonumber\\
&= \frac{S^w-1}{ln\ S}\label{eqn:exactLw}\\
\nonumber\\
&= \frac{(1-P)^{Ew}-1}{ln\ (1-P)^{E}}\nonumber
\end{align}

To derive a simple approximation for $\Lww$, recall from Equation~\ref{eqn:Nt-approxB} that $N(w)\approx B/(EP)$.  It is immediately apparent from Equation~\ref{eqn:NBL} that:
\begin{align}
\Lww\approx \frac{1}{EP}\forWsmallP
\label{eqn:Lapprox}
\end{align}

Finally, the ratio of Equation~\ref{eqn:LDFiftyEP} to Equation~\ref{eqn:Lapprox} is noteworthy:
$${LD_{50}}\ /\ {\Lww} \approx ln\ 2 \approx 0.7\forWsmallP$$
%=================================================
\mathsubsection{Maximum $\Lww$}

To find the $w$ where $\Lww$ is maximal, we set the derivative of the definition of $\Lww$ (from Equation~\ref {eqn:exactLw}) to zero:
\begin{align}
0 &= \frac{d[\frac{S^w-1}{ln\ S}]}{dw}\nonumber\\
\nonumber\\
&= \frac{1}{ln\ S}\ S^w\ ln\ S\nonumber\\
\nonumber\\
&= S^w\nonumber\nonumber
\end{align}
Given $0<S<1$, then $S^w=0$ at $w= \infty$.  So, using Equations~\ref{eqn:exactLw} and \ref{eqn:PlnP}:
\begin{align}
\Lww_{max} = \overline{L(\infty)} &= \frac{S^\infty-1}{ln\ S}\nonumber\\
&= \frac{-1}{E\ ln(1-P)}\nonumber\\
&= \frac{1}{EP}\ \text{as}\ P\rightarrow0\nonumber
\end{align}
\def\brackfor{[\textrm{for}\ }

\noindent Seeking to show $\Lww_{max}<1/(EP)$ for all $P$, we begin by observing:
\begin{equation}
x - \frac{1}{2}x^2 + \frac{1}{3}x^3 ... < x\hskip 5ex\brackfor x<0]\nonumber
\end{equation}

\noindent The Mercator series is:
\begin{equation}
x - \frac{1}{2}x^2 + \frac{1}{3}x^3 ... = ln\ (1+x)\hskip 5ex\brackfor -1<x\le 1]\nonumber
\end{equation}

\noindent We combine the two preceding formulae into an inequality, then set $x=-P$:
\begin{align}
ln\ (1+x) &< x\nonumber&&\brackfor -1<x<0]
\\
ln\ (1-P) &< -P\nonumber&&\brackfor -1<-P<0]
\end{align}

\noindent Substituting back into the definitions of $\Lww_{max}$ gives, for $0 < P < 1$:
\begin{equation}
\Lww_{max} = \overline{L(\infty)} = \frac{-1}{E\ ln(1-P)} =  \bigg(\lim\limits_{P\rightarrow 0} \frac{1}{EP}\bigg)< \frac{1}{EP}
\label{eqn:maxLwwLimit}
\end{equation}

%=================================================
\mathsubsection{Model of $E$ Growing Over Time}

\revised{We wish to model $E$ growing over time and apply the model to time spans that do not cause exponential explosion.}

\ 

\noindent\revised{First, recall Equation~\ref{eqn:defCy}, the basis for the baseline ``Constant-$E$'' model:}

\begin{equation} 
C(y) = (1-P)^{(Ey)}\tag{\ref{eqn:defCy}}
\end{equation}

\noindent\revised{The exponential term $Ey$ has units entity-years, and signifies the total exposure of the civilization to destruction events. Renaming $E$ to $E_0$ to reinforce its constant nature in the Constant-$E$ model, we can write:}

\begin{equation} 
Exposure[Constant Model]_y = E_0\thinspace y
\label{eqn:constantExposure}
\end{equation}

\noindent\revised{Similarly, we can calculate exposure for a model in which $E$ grows with time.  Equation~\ref{eqn:defEgrowth} defined $E_y$ as growing from an initial value of $E_0$ at an annual rate of $r$ over a period of $y$ years, with the growth continuously compounded:}

\begin{equation} 
E_y = E_0\thinspace e^{ry}\tag{\ref{eqn:defEgrowth}}
\end{equation}

\noindent\revised{In this ``Growing-$E$'' model, the civilization's exposure to destruction events is:}

\begin{align} 
\nonumber Exposure[GrowthModel]_{y} &= \int_{t=0}^y E_0\ e^{rt}\ dt
\\
\nonumber &= E_0\ \bigg(\frac{e^{rt}}{r}\bigg\rvert_{t=0}^y \bigg)
\\
&= E_0\ \frac{e^{ry}-1}{r}\label{eqn:growthExposure}
\end{align}

\ 

\noindent\revised{We can equate the two exposures from the right-hand-sides of Equations~\ref{eqn:constantExposure} and \ref{eqn:growthExposure}, taking care to distinguish the two different $y$:}

\begin{align*} 
E_0\thinspace y_1 = E_0\ \frac{e^{ry_2}-1}{r}
\end{align*}

\noindent\revised{This equation says that a civilization's destruction-exposure after $y_1$ years as calculated by the constant model, equals the exposure after $y_2$ years as calculated by the growth model.}

\ 

\noindent\revised{Continuing, we can cancel the $E_0$ terms and express $y_2$ in terms of $y_1$:}

\begin{align} 
\nonumber y_1 &= \frac{e^{ry_2}-1}{r}
\nonumber \\
\nonumber \\
y_2 &= \frac{ln\ (1 + ry_1)}{r}\label{eqn:yOneyTwo}
\end{align}

\noindent\revised{Analytically, Equation~\ref{eqn:yOneyTwo} provides a shortcut for converting results from the constant-$E$ model to the growing-$E$ model.  For example, we can define the $LD_{50}$ for the growing-$E$ model as:}

\begin{equation}
LD_{50}[G] = \frac{ln\ (1 + r\thinspace LD_{50}[C])}{r}
\label{eqn:LDFiftyEq}
\end{equation}

\noindent\revised{where the [G] and [C] indicate the growing-$E$ model and constant-$E$ model, respectively.  See Figure~\ref{fig:growthmodel}.}

\ 

\noindent\revised{Although Equation~\ref{eqn:yOneyTwo} does not have an explicit exponential term, it must still be applied carefully because it implicitly assumes that the number of entities can grow exponentially without limit, per Equation~\ref{eqn:defEgrowth}. Alternative growing-$E$ models may be derived, e.g.\ using linear growth.}

%%%%%%%%%%%%%%%%%%%%%%%%%%%%%%%%%%%%%%%%%%%%%%%%%%%%%%%%%%

\vfill\break
\nocite{mpmath}
\bibliography{fermi-math}{}
\bigskip

% The format that I like for bibliographies.
% IEEE from: https://stackoverflow.com/questions/144639/how-to-order-citations-by-appearance-using-bibtex
%\bibliographystyle{ieeetr}

% The Harvard style that IJA wants
\bibliographystyle{agsm}

\end{document}